\documentclass[letterpaper, 10 pt, conference]{ieeeconf}  
\usepackage[letterpaper, left=0.97in, right=0.97in, bottom=0.80in, top=0.77in]{geometry} 

\IEEEoverridecommandlockouts                              

\usepackage{graphics} 
\usepackage{xcolor}
\usepackage[font=normalsize]{caption}
\usepackage{epsfig} 
\usepackage{amsmath} 
\usepackage{amssymb}  
\usepackage[caption=false,font=footnotesize]{subfig}
\usepackage[font=normalsize]{caption}
\usepackage{algorithm}
\usepackage{algpseudocode}
\usepackage{mathabx}
\usepackage{bm}
\usepackage{dsfont}
\usepackage[export]{adjustbox}

\usepackage{comment}

\newcommand{\R}{\mathbb{R}}
\newcommand{\Tau}{\mathcal{T}}
\newcommand{\B}{\mathcal{B}}
\renewcommand{\baselinestretch}{0.95}

\title{\LARGE \bf
Time Shift Governor for Constrained Control of Spacecraft Orbit and Attitude Relative Motion in Bicircular Restricted Four-Body Problem
}

\author{Taehyeun Kim, Ilya Kolmanovsky, and Anouck Girard 
\thanks{Taehyeun Kim, Ilya Kolmanovsky, and Anouck Girard are with the Department of Aerospace Engineering, University of Michigan, Ann Arbor, 48109 MI, USA. {\tt\small \{taehyeun, ilya, anouck\}@umich.edu.}\newline
\indent This research is supported by the Air Force Office of Scientific Research Grant number FA9550-20-1-0385 and FA9550-23-1-0678.}
}

\begin{document}

\maketitle
\thispagestyle{empty}
\pagestyle{empty}

\begin{abstract}
This paper considers constrained spacecraft rendezvous and docking (RVD) in the setting of the Bicircular Restricted Four-Body Problem (BCR4BP), while accounting for attitude dynamics. We consider Line of Sight (LoS) cone constraints, thrust limits, thrust direction limits, and approach velocity constraints during RVD missions in a near rectilinear halo orbit (NRHO) in the Sun-Earth-Moon system. To enforce the constraints, the Time Shift Governor (TSG), which uses a time-shifted Chief spacecraft trajectory as a target reference for the Deputy spacecraft, is employed. The time shift is gradually reduced to zero so that the virtual target gradually evolves towards the Chief spacecraft as time goes by, and the RVD mission objective can be achieved. Numerical simulation results are reported to validate the proposed control method.
\end{abstract}
\vspace{-0.05 in}

\section{INTRODUCTION}\label{sec:intro}  
\vspace{-0.05 in}
The recent Global Exploration Roadmap published by the International Space Exploration Coordination Group envisions the development and operation of an outpost in the cislunar space, called the Lunar Orbital Platform-Gateway (LOP-G). Autonomous rendezvous and docking (RVD) technology is an integral part of the LOP-G, enabling supply delivery, on-orbit maintenance, large-scale structure assembly, and lunar sample return.

A near rectilinear halo orbit (NRHO) family around the $L_2$ Lagrange point in the Earth-Moon system has been proposed as a destination for the Lunar Gateway~\cite{williams2017targeting}. In particular, the 9:2 southern $L_2$ NRHO in the Earth-Moon system, which is considered in this work, has been chosen as the target orbit for the Gateway, as it provides low orbit maintenance cost, favorable communication opportunities, and safe power supply, which come from eclipse avoidance~\cite{guzzetti2017stationkeeping}.

The coupled translational and rotational dynamics have been considered for RVD in near-Earth orbits~\cite{kasiri2023coupled} and in the circular restricted three-body problem setting~\cite{bucchioni2021rendezvous}. Colagrossi and Lavagna \cite{colagrossi2017preliminary} have studied a coupled orbit-attitude dynamical model that addresses the effects of large structural flexibility, considering the Sun's gravitational effect and solar radiation pressure.

In this paper, we consider the control of coupled orbit and attitude dynamics in the Bicircular Restricted Four-Body Problem (BCR4BP) setting. The Circular Restricted Three-Body Problem (CR3BP) is commonly considered in the preliminary cislunar trajectory design. The BCR4BP model is an extension of the Earth-Moon CR3BP, as it accounts for the fourth body gravitational influence, the Sun gravity effect - disregarded by the CR3BP model. The BCR4BP model sustains important geometric properties of target orbits in the CR3BP, such as perilune and apolune radii and eclipse avoidance. This coherence makes it a valuable substitute for NRHOs in the CR3BP, allowing for realistic dynamical simulations.
By using the BCR4BP to leverage the dynamical equivalents of the 9:2 NRHO in the CR3BP framework originally intended for the Lunar Gateway, our goal is to demonstrate the effectiveness of our proposed control scheme within the modeled cislunar environment, thereby reducing the gap between the model, used by mission designers, and the actual dynamics of the system. 

Various control schemes have been considered for RVD missions with coupled orbit and attitude dynamics. A linear quadratic controller~\cite{moon2016quaternion} solving a Two-point Boundary Value Problem has been employed for spacecraft RVD around Earth orbits; its design was based on the Clohessy-Wiltshire-Hill equations for translational motion, and a quaternion-based PD attitude controller tracks the desired orientation of the spacecraft. 
For spacecraft RVD in near-rectilinear halo orbits, a PID controller~\cite{bucchioni2021rendezvous} and a nonlinear control algorithm~\cite{muralidharan2023rendezvous} that relies on an Interior Point Optimizer (IPOPT) have also been considered.

The TSG is a variant of a parameter governor~\cite{kolmanovsky2006parameter} that adjusts parameters in the nominal control law to satisfy pointwise-in-time state and control constraints at a low computational cost. The TSG adjusts the time shift along the reference trajectory to satisfy constraints and achieve convergence. The TSG has been previously applied to spacecraft formation control in circular Earth orbits~\cite{2016gregory}, RVD in elliptic Earth orbits~\cite{kim2024time}, and RVD in Halo orbits in the CR3BP setting~\cite{kim2023time}, where the attitude dynamics of the spacecraft were not considered. In this paper, we extend the TSG for halo orbit RVD missions in the BCR4BP setting, incorporating coupled translational and attitude dynamics.

The rest of the paper is organized as follows. Section~\ref{sec:problem} describes the problem formulation, including models for spacecraft translational dynamics in the BCR4BP setting and rotational dynamics. Section~\ref{sec:control} outlines the nominal control system design and constraints that are addressed during the RVD mission. The TSG is discussed in Section~\ref{sec:TSG}. Section~\ref{sec:results} provides numerical simulation results demonstrating the capability of the TSG to enforce constraints. Lastly, Section~\ref{sec:conclusion} presents conclusions and future research directions.

\vspace{-0.05 in}

\section{PROBLEM FORMULATION} \label{sec:problem}
\vspace{-0.05 in}
A spacecraft rendezvous and docking mission (RVD) is considered in a near rectilinear halo orbit (NRHO) from the perspective of satisfying various mission-specific constraints. During the RVD mission, we assume that the primary spacecraft, named Chief, is located further behind the secondary spacecraft, named Deputy, along the orbital track at the initial time instant. A reverse situation when the Deputy is located behind the Chief is addressed similarly. Subscripts $c$ and $d$ designate the Chief spacecraft and the Deputy spacecraft, respectively.\vspace{-0.05 in}

\subsection{Coordinate Systems} \label{secFrame} 
\begin{figure}[htbp!]
     \includegraphics[width=0.77\linewidth, center]{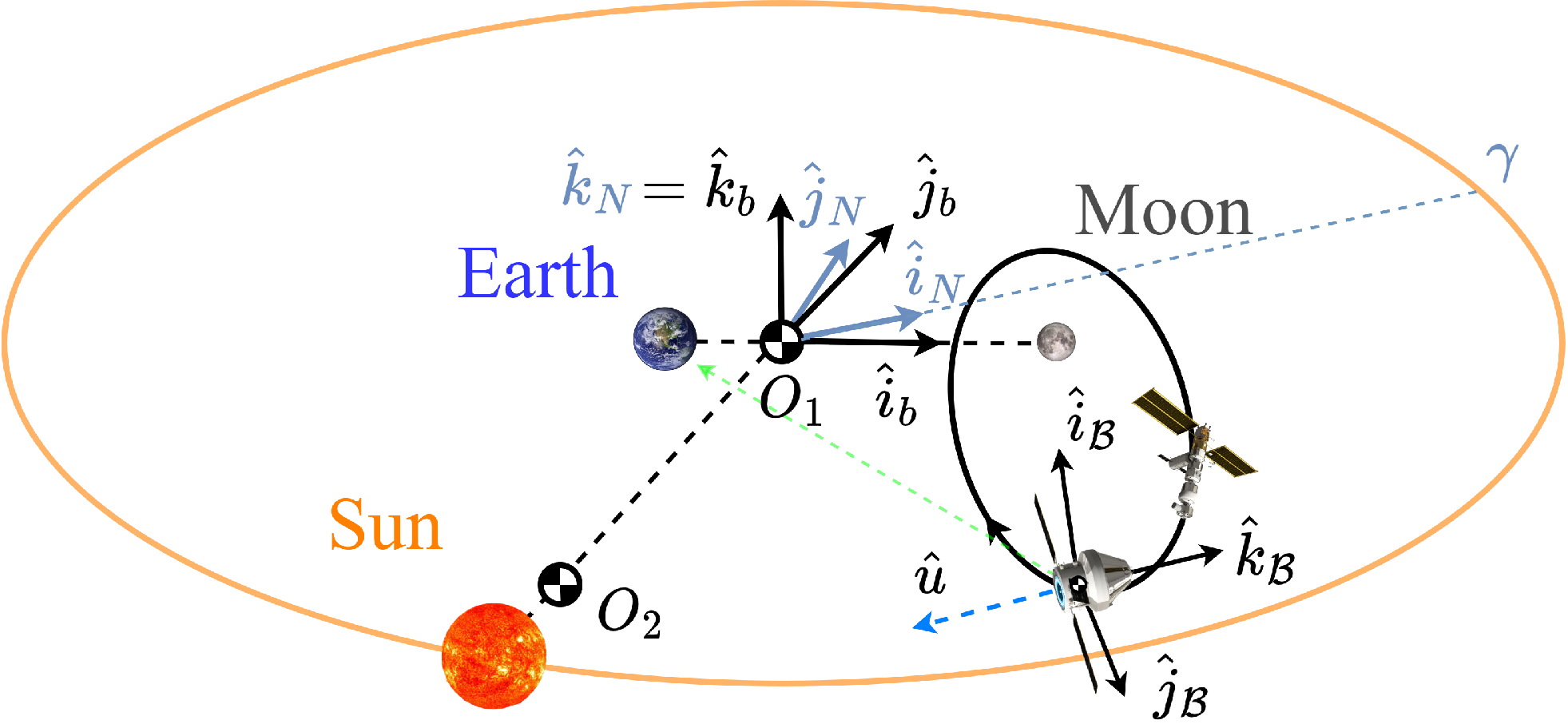}
\caption{Barycentric frame $b$ and Body-fixed frame $\B$ in the Sun-Earth-Moon system.} \vspace{-0.1 in}
\label{figFrame}    
\end{figure}

\vspace{-0.05 in}
Three different frames are used here: an inertial frame $N$, the barycentric frame $b$, and the body-fixed frame $\B$. The spacecraft dynamics are first written in the barycentric frame, assumed rotating with respect to the inertial frame. This barycentric frame is defined by $b: \{ \bm{O}_{1}, \bm{\hat{i}}_{b}, \bm{\hat{j}}_{b}, \bm{\hat{k}}_{b} \}$ where $\bm{O}_{1}$ is the center of mass of the Earth-Moon system, $\bm{\hat{i}}_{b}$ points in the direction from the Earth to the Moon, $\bm{\hat{k}}_{b}$ is aligned with the Earth-Moon system angular momentum vector relative to the Sun, and $\bm{\hat{j}}_{b}$ completes the right-handed system, as shown in Figure~\ref{figFrame}. The inertial frame, $N=\{ \bm{O}_{1}, \bm{\hat{i}}_{N}, \bm{\hat{j}}_{N}, \bm{\hat{k}}_{N} \}$, has the same origin and $\bm{\hat{k}}_{N}$ as the barycentric frame.
The body-fixed frame is defined by $\B: \{ \bm{O}_{\B}, \bm{\hat{i}}_{\B}, \bm{\hat{j}}_{\B}, \bm{\hat{k}}_{\B} \}$ where $\bm{O}_{\B}$ is the center of mass of the Deputy spacecraft, the body frame $\B$ is chosen such that the inertia expressed in $\B$ is diagonal and a single thruster aligns with $-\bm{\hat{k}}_{\B}$, as illustrated in Figure~\ref{figFrame}.\vspace{-0.05 in}

\subsection{Coupled Orbit and Attitude Dynamics} \vspace{-0.05 in}
The spacecraft equations of motion can be expressed in the barycentric frame as \vspace{-0.05 in}
\begin{equation}
    \dot{\bar{X}}_{i}(\tau) = \bar{f}\big(\tau, \bar{X}_{i}(\tau), u_{i}(\tau), M_{i}(\tau) \big),
    \label{eqDyn}
\end{equation} 
where $\bar{X}_i \in \R^{12}$, $i \in \{c,d\}$ incorporates the position, velocity, attitude, and angular velocity of the spacecraft, $u_i\in \R^3$, $i \in \{c,d\}$ is the translational control input (thrust-induced acceleration) to the spacecraft, $M_i \in \R^3$ is the rotational control input (control moment) applied to the spacecraft by the attitude control system (e.g., reaction wheels or CMGs), and $\tau \in \mathbb{R}_{\geq 0}$ denotes time. We assume that the docking port of the Chief spacecraft automatically aligns with its velocity direction, and focus on controlling the Deputy spacecraft to close proximity of the Chief spacecraft without violating mission constraints. We assume that the final docking procedure, e.g., with the assistance of a robotic arm or by engaging the terminal docking controller, takes place once the Deputy spacecraft is in close proximity of the Chief spacecraft, in the same halo orbital track and in the correct orientation to initiate the actual docking procedure. \vspace{-0.05 in}

\subsection{Bicircular Restricted Four-Body Problem} \vspace{-0.05 in}
The translational motion of the spacecraft is modeled using the Bicircular Restricted Four-Body Problem (BCR4BP) formulation. The BCR4BP is an extension of the Circular Restricted Three-Body Problem (CR3BP) to account for the gravitational influence of the third massive body~\cite{jorba2018two}. In the CR3BP setting, we assume that the two primary celestial bodies, the Earth and the Moon, are point masses in circular orbits around their shared barycenter, referred to as $O_{1}$ in Figure~\ref{figFrame}~\cite{cr3bp_nd}. In the BCR4BP, additionally, the Sun and $O_{1}$ are assumed to follow a circular orbit centered at the barycenter of the Earth-Moon-Sun system, referred to as $O_{2}$ in Figure~\ref{figFrame}, sharing the same orbital plane with the Earth-Moon plane of motion, as illustrated in Figure~\ref{figFrame}. Note that the units of distance and time are normalized, respectively, by the Earth-Moon distance and the mean motion of the Moon. 
 
The equations of motion for the BCR4BP are given in non-dimensional form as \vspace{-0.05 in}
\begin{equation} \label{eq:crtbpx}
\begin{aligned}
    \ddot{x} &= 2\dot{y} + \frac{\partial \tt U}{\partial x} + \frac{\partial \Gamma}{\partial x} + u_{x}, \\
    \ddot{y} &= -2\dot{x} + \frac{\partial \tt U}{\partial y} + \frac{\partial \Gamma}{\partial y} + u_{y}, \\
    \ddot{z} &=  \frac{\partial \tt U}{\partial z} + \frac{\partial \Gamma}{\partial z} + u_{z},
\end{aligned}
\end{equation} 
where $x,y,z$ are the coordinates in the barycentric frame; $\tt U$ and $\Gamma$ are the pseudo-potentials stemming from the Earth-Moon system and the Sun, respectively. These pseudo-potentials are defined as~\cite{jorba2018two,cr3bp_nd} \vspace{-0.05 in}
\begin{equation}
    \begin{aligned}
        \text{$\tt U$}& = \frac{1}{2}(x^2 + y^2) + \frac{1-\mu}{ r_{sc/ \oplus} } + \frac{\mu}{ r_{sc/ \leftmoon} },    \\
        \Gamma &= \frac{\mu_{\odot}}{r_{sc/\odot}} - \frac{\mu_{\odot}}{a^{3}_{\odot}}(x_{\odot} x + y_{\odot} y + z_{\odot} z),
    \end{aligned}    
\end{equation}
where \vspace{-0.05 in}
\begin{equation}
    \begin{aligned}
        r_{sc/ \oplus} &= \sqrt{(x + \mu)^2 + y^2 + z^2}, \\
    r_{sc/ \leftmoon} &= \sqrt{(x + \mu - 1 )^2 + y^2 + z^2}, \\
    r_{sc/\odot} &= \sqrt{(x - x_{\odot} )^2 + (y - y_{\odot} )^2 + (z - z_{\odot} )^2},
    \end{aligned}
\end{equation}  
and where $\mu$ represents the mass ratio of the Moon to the total mass of the Earth-Moon system, $m_{\oplus}$ is the mass of the Earth, and $m_{\leftmoon}$ is the mass of the Moon. Additionally, we use $\mu_{\odot}$ to denote the ratio of the Sun's mass to the total mass Earth-Moon system, and $a_{\odot}$ represents the distance between the Sun and the Earth-Moon barycenter. Note the spacecraft's state and control input are expressed in the barycentric frame $b$ in Figure~\ref{figFrame}. The variables, $x_{\odot},\; y_{\odot},$ and $z_{\odot}$, are the components of the position vector of the Sun relative to $O_{1}$, expressed in the Earth-Moon barycentric frame,  \vspace{-0.05 in}
\begin{equation}
    \begin{bmatrix}
        x_{\odot}\\ y_{\odot} \\ z_{\odot}
    \end{bmatrix} = a_{\odot} 
    \begin{bmatrix}
        \cos(\theta_{\odot}(\tau)) \\
        \sin(\theta_{\odot}(\tau)) \\
        0
    \end{bmatrix} = a_{\odot} 
    \begin{bmatrix}
        \cos(\omega_{\odot} \tau + \theta_{0}) \\
        \sin(\omega_{\odot} \tau + \theta_{0}) \\
        0
    \end{bmatrix},
\end{equation}
where $\omega_{\odot}=-0.9252$ and $\theta_{0}$ represent the Sun's angular velocity and an initial angle of the Sun, as measured from the $\hat{i}_{b}$ axis in the Earth-Moon barycentric frame. The Sun angle, $\theta_{\odot}$, is a function of the nondimensional time $\tau$ and also depends on $\omega_{\odot}$ and $\theta_{0}$. Note that by incorporating the gravitational influence of the Sun, \eqref{eqDyn} becomes a non-autonomous system because the position of the Sun changes as a function of time. We note that the BCR4BP model and the subsequent TSG design can be extended to the ECR4BP model, which, in addition, accounts for the eccentricity of the Moon and Earth orbits and depends on the true anomaly of Moon orbital motion around the Earth. \vspace{-0.1 in}

\subsection{Attitude Kinematics and Dynamics} \vspace{-0.05 in}
We use modified Rodrigues parameters (MRP) to represent the Deputy spacecraft's attitude. This parameterization comes from a stereographic projection of the quaternion unit sphere onto the MRP hyperplane. The vector of MRPs $\sigma$ can be expressed in terms of the Euler parameters $\beta$ or the principal rotation elements $(\hat{e},\Phi)$ as $\sigma_{i}=\tan\frac{\Phi}{4}\hat{e}=\frac{\beta_i}{1+\beta_0},\; i=\{ 1,2,3 \}$
where $\hat{e}$ and $\Phi$ are the principal axis unit vector and rotation angle, respectively, and the Euler parameters are defined by $\beta_0=\cos(\Phi/2)$ and $\beta_{i} = {e}_{i} \sin(\Phi/2),\; i=1,2,3$.
The kinematic equations of motion using the MRPs are  \vspace{-0.05 in}
\begin{equation}
    \dot{\sigma}_{\B/b} = \frac{1}{4}\bigg[ ( 1 - \sigma^{\sf T}_{\B/b} \sigma_{\B/b} )[I] + 2[\Tilde{\sigma}_{\B/b}] + 2 \sigma_{\B/b}\sigma^{\sf T}_{\B/b} \bigg] {}^{\B}\omega_{\B/b},
\end{equation}
where $[I]$ is a $3\times 3$ identity matrix and ${\sigma}_{\B/b}$ is the attitude of the body-fixed frame $\B$ with respect to the barycentric frame $b$ represented by the MRP. ${}^{\B}{\omega}_{\B/b}$ is the angular velocity of the body-fixed frame with respect to the barycentric frame, expressed in the body-fixed frame. The tilde operator, $[\Tilde{\cdot}]$, is a skew-symmetric matrix defined by  \vspace{-0.05 in}
\begin{equation}
    [\Tilde{\omega}] = \begin{bmatrix}
        0&-\omega_3 & \omega_2\\
        \omega_3&0 & -\omega_1\\
        -\omega_2&\omega_1 &0
    \end{bmatrix},\; \text{for } \omega=[\omega_1,\; \omega_2,\; \omega_3]^{\sf T},
\end{equation}
and $\vee:\R^{3\times 3} \to \R^{3}$ is the inverse of the tilde operator, i.e., $\omega = [\Tilde{\omega}]^{\vee}$.

The Euler rotational equations of motions are: $\dot{\omega}_{\B/b}=[{\tt I_{sc}}]^{-1}\big[-[\Tilde{\omega}_{\B/b}][{\tt I_{sc}}]\omega_{\B/b} + M \big]$, where $[{\tt I_{sc}}]$ is the moment of inertia of the Deputy spacecraft. Note that $\omega_{b/N}=\hat{k}_{b}$ is constant based on the assumptions in the setting of the BCR4BP, resulting in $\dot{\omega}_{b/N}=0$, i.e., $\dot{\omega}_{\B/N}=\dot{\omega}_{\B/b}$.

\vspace{-0.05 in}

\section{NOMINAL CONTROLLER DESIGN} \label{sec:control}
\vspace{-0.05 in}
We design the nominal controller to track the translational motion of the target and to control the rotational motion of the Deputy spacecraft with the desired thrust direction. \vspace{-0.1 in}

\subsection{Averaged-in-time LQR} \vspace{-0.05 in}
The primary goal of the nominal controller is to track a translational state reference, including position and velocity, that corresponds to the Chief spacecraft in the reference orbit or to a virtual target, determined by a time-shifted state of the Chief spacecraft along the orbital track. We employ the averaged-in-time linear-quadratic regulator (ALQR))~\cite{kalabic2015station} as our nominal controller.

Let $(A,B)=\frac{1}{N}\sum^{N-1}_{k=0}(A_k,B_k)$ be the pair of the averaged dynamics-input matrices, where the pair $(A_k,B_k)$ denotes a linearized translational dynamics-input pair at time $\tau_k$ and $N$ is the number of equidistant time instants over a single Chief spacecraft orbit period so that  \vspace{-0.05 in}
\begin{equation}
    \begin{split}
        \delta \dot{X}_{d} &= \bigg[ \frac{\partial f}{\partial X_{d}} \big(\tau_k, X_{v}, 0 \big) \bigg] \delta X_{d} + \bigg[ \frac{\partial f}{\partial u_{d}}\big(\tau_k, X_{v}, 0 \big) \bigg] \delta u_{d}, \\
        &= A_k \delta X_{d} + B_k \delta u_{d}, 
    \end{split} 
    \label{eqLinearizedDyn2Target}
\end{equation} 
where $\delta X_{d} = X_{d}(\tau) - X_{v}(\tau), \delta u = u_{d}(\tau) - 0, $ and $X_{v}(\tau)$ denotes the virtual target state for the Deputy spacecraft. We assume that the Chief spacecraft operates in an NRHO and follows an unforced periodic natural motion trajectory (i.e., $u_{c}(\tau) = 0, \forall \tau \in \mathbb{R}_{\ge 0}$), while the Deputy tracks the virtual target using the feedback law. With selected symmetric positive-definite cost matrices $Q\in \R^{6\times 6}$ and $\mathcal{R}\in \R^{3\times 3}$, the ALQR is a solution to the following optimal control problem, \vspace{-0.05 in}
\begin{equation} \label{eqOptProb_translational}
\begin{aligned} 
\min \int^{\infty}_{0} \delta X^{\sf T}_{d}(\tau) Q \delta X^{\sf T}_{d}(\tau) + \delta u^{\sf T}_{d}(\tau) \mathcal{R} \delta u^{\sf T}_{d}(\tau) d\tau, \\ 
\text{subject to }\quad \quad \delta \dot{X}_{d} =A \delta X_{d} + B \delta u_{d}.
\end{aligned}
\end{equation}
The solution to \eqref{eqOptProb_translational} is then a feedback control law that provides the desired thrust $u_{d}(\tau),$ 
\begin{equation} \label{eqNominalCtrl}
    u_{d}(\tau) = K \delta X_d(\tau),\; \hat{u}_{d} = {u}_{d}/ \| {u}_{d} \|,
\end{equation}
where $K =-\mathcal{R}^{-1}B^{\sf T}P,$ and $P$ is the positive semi-definite solution to the algebraic Riccati equation, $0=A^{\sf T}P + PA - PB\mathcal{R}^{-1}B^{\sf T}P + Q$. Note that TSG is applicable to other nominal controllers, such as LQR with gain re-computed along the orbit, as long as the nominal controller is (locally) stabilizing. Such nominal controllers have to ensure (local) uniform asymptotic stability of the unforced trajectory $X_v$, for the Deputy spacecraft dynamics, i.e., $X_d(\tau) \to X_v(\tau)$ as $\tau \to \infty$.\vspace{-0.05 in}

\subsection{Geometric Tracking Control} \vspace{-0.05 in}
The objective of the nominal attitude controller is to align the actual thrust direction, $\hat{u}$ in Figure~\ref{figFrame}, with the desired thrust direction, $\hat{u}_{d}$. Since the spacecraft has a single thruster acting along $\hat{k}_{\B}$, the actual thrust is then \vspace{-0.05 in}
\begin{equation} \label{eqActualCtrl}
    u(\tau)=-|u_{d}(\tau)| \hat{k}_{\B}.
\end{equation}
We use a geometric tracking control law that ensures exponential stability at the zero equilibrium of the attitude tracking errors if the initial attitude error is less than $180^{\circ}$, see~\cite{lee2010geometric}. 
The desired attitude $R$ of the spacecraft is defined as \vspace{-0.05 in}
\begin{equation} \label{eqDesiredAtt}
    [Rb]=\begin{bmatrix}
        (\hat{r}_{\oplus/d} \times \hat{u}_{d})^{\sf T} \\ 
        (-\hat{u}_{d}\times (\hat{r}_{\oplus/d} \times \hat{u}_{d}))^{\sf T} \\
        - \hat{u}^{\sf T}_{d} 
    \end{bmatrix},
\end{equation}
 \begin{equation} \label{eqDesiredAttRate}
     \dot{[Rb]}=\begin{bmatrix}
         \dot{\hat{r}}_{\oplus/d} \times \hat{u}_{d} + {\hat{r}}_{\oplus/d} \times \dot{\hat{u}}_{d} \\ 
         [(\hat{r}_{\oplus/d} \times \hat{u}_{d}) \times \dot{\hat{u}}_{d} \quad \cdots \qquad \\ \quad + (\dot{\hat{r}}_{\oplus/d} \times \hat{u}_{d} + \hat{r}_{\oplus/d} \times \dot{\hat{u}}_{d}) \times \hat{u}_{d}] \\
         - \dot{\hat{u}}_{d} 
     \end{bmatrix},
 \end{equation} 
where \vspace{-0.05 in}
\begin{equation}
    \begin{aligned}
        \vec{r}_{\oplus/d}&=p(X_{\oplus}-X_{d}),&\quad \hat{r}_{\oplus/d}&=\vec{r}_{\oplus/d}\|\vec{r}_{\oplus/d}\|, \\
        \quad \dot{\vec{r}}_{\oplus/d}&=v({X}_{\oplus}-{X}_{d}),&\quad \dot{\hat{r}}_{\oplus/d}&=\dot{\vec{r}}_{\oplus/d} / \|\dot{\vec{r}}_{\oplus/d}\|, \\        
        \quad \dot{\vec{u}}_{d}&=K(\dot{X}_{d}-\dot{X}_{v}),&\quad \dot{\hat{u}}_{d}&=\dot{\vec{u}}_{d} / \|\dot{\vec{u}}_{d}\| .
    \end{aligned}
\end{equation}
In \eqref{eqDesiredAtt}, $[Rb]\in \R^{3\times 3}$ denotes the direction cosine matrix (DCM) of the desired reference frame $R$ with respect to the barycentric frame $b$. The DCM $[Rb]$ in \eqref{eqDesiredAtt} is made by stacking three physical vectors, expressed in the barycentric frame $b$. Remark that a DCM $[AC]$ denotes a matrix that maps physical vectors in the $C$ frame into $A$ frame vectors, where $A=\{ \hat{a}_{i}, \hat{a}_{j}, \hat{a}_{k} \}$ and $C=\{ \hat{c}_{i}, \hat{c}_{j}, \hat{c}_{k} \}$ are two arbitrary frames, and the entries of the DCM are $[AC]_{ij} = \cos \alpha_{ij} = \hat{a}_{i} \cdot \hat{c}_{j}$. With this notation, the transpose of a DCM can be expressed by changing the order of letters in the DCM, i.e., $[CA]=[AC]^{\sf T}$. The kinematic differential equation for the DCM is given by $\dot{[Rb]} = -[{}^{R}\Tilde{\omega}_{R/b}] [Rb],$ where $\dot{[Rb]}$ is the time derivative of $[Rb]$ in \eqref{eqDesiredAtt} and the left superscript indicates the coordinate system in which the angular velocity is expressed. The angular velocity of the reference frame can be obtained as \vspace{-0.05 in}
\begin{equation} \label{eqAngVel}
    {}^{R}{\omega}_{R/b}= -(\dot{[Rb]}[Rb]^{\sf T})^{\vee }.
\end{equation}
The attitude tracking error, $e_{[C]}$, is defined by $e_{[C]} = \frac{1}{2} ([Rb][b \B]-[\B b][bR])^{\vee},$ where $\vee : \R^{3\times 3} \to \R^3$ maps a DCM to a vector, i.e., represents the inverse of the tilde operator, $e_{\omega} = {}^{\B}\omega_{\B/b} - [\B b][bR]{}^{R}\omega_{R/b}$.
The time rate of change of the angular velocity is \eqref{eqAngVel}: \vspace{-0.05 in}
\begin{equation} \label{eqAngVelRate}
    {}^{b}\frac{d}{dt}\bigg( \omega_{R/b} \bigg) = {}^{R}\frac{d}{dt}\bigg( {}^{R} \omega_{R/b} \bigg) 
    = \Bigg{.}^{R}\begin{bmatrix}
        \dot{\omega}_{R/b,1} \\ \dot{\omega}_{R/b,2} \\ \dot{\omega}_{R/b,3}
    \end{bmatrix}
\end{equation} 
and the feedback law for the control moment is given by
\begin{equation} \label{eqAttControl} 
    \begin{aligned}
        M &= -k_{P} e_{[C]} - k_{D} e_{\omega} + {}^{\B}\omega_{\B/b} \times [{\tt I_{sc}}] {}^{\B}\omega_{\B/b} \\
        &-[{\tt I_{sc}}] \bigg( [{}^{\B}\Tilde{\omega}_{\B/b}][\B b][bR]{}^{R}{\omega}_{R/b} - [\B b][bR]{}^{R}{\dot{\omega}}_{R/b} \bigg),
    \end{aligned}
\end{equation}
where $k_P, k_D,$ and $[{\tt I_{sc}}]$ denote the P gain, D gain, and moment of inertia tensor of the spacecraft expressed in $\B$, respectively. If we assume that the angular velocity with respect to the barycentric frame changes sufficiently slowly, we can omit the last term, i.e., ${}^{\B}{\dot{\omega}}_{R/b}$, in \eqref{eqAttControl}.\vspace{-0.1 in}

\subsection{Constraints} \label{sec:constraints} \vspace{-0.05 in}
The Deputy spacecraft performing the rendezvous and docking (RVD) mission faces various constraints. We consider four types of constraints to demonstrate the effectiveness of our method: a line of sight (LoS) cone angle constraint, a limit on magnitude of thrust, a limit on thrust direction, and a relative velocity constraint in the proximity of the Chief spacecraft.

While approaching the Chief, the Deputy has to operate within a prescribed  Line of Sight (LoS) cone, which is defined by a LoS half-cone angle $\alpha$ as \vspace{-0.05 in}
\begin{equation}
    \begin{split}
        h_{1} &= -  v(X_{c})^{\sf T} p(X_{d} - X_{c}) \\
        & \quad + \cos(\alpha) \big\| v (X_{c}) \big\|  \big\| p( X_{d} - X_{c}) \big\| \leq 0, \\
    \end{split}
    \label{eqDefConeConst}
\end{equation}
where $p(\cdot):\R^{6} \to \R^{3}$ draws the position vector and $v(\cdot):\R^{6} \to \R^{3}$ draws the velocity vector, corresponding to the full state $X$.

The thrust limit constraint is given by \vspace{-0.05 in}
 \begin{equation}
    h_{2} = | u_{d} | - u_{\max} \leq 0,
    \label{eqDefCtrlConst}
\end{equation}
where $u_{\max}$ denotes the maximum magnitude of the control input. Instead of managing \eqref{eqDefCtrlConst} by TSG, the saturation function is used to enforce \eqref{eqDefCtrlConst} as a part of the nominal controller as this typically leads to a faster response~\cite{cotorruelo2019output}. The controller then takes the form \vspace{-0.05 in}
\begin{equation}
    u_{d}(\tau) : = \begin{cases} u_{d}(\tau), &\text{if } |u_{d}(\tau) | \leq u_{\max}, \\
    u_{\max} \hat{u}_{d}(\tau), &\text{if } |u_{d}(\tau) | > u_{\max}. \end{cases}
    \label{eqSat}
\end{equation}
The TSG takes into account the saturation of the control input in its prediction model based on \eqref{eqSat}.

The angle between the actual Deputy's thrust direction and the desired direction is restricted by the maximum angle difference $\eta$, resulting in \vspace{-0.05 in}
\begin{equation} \label{eqDefCtrlDirectionConst}
    h_{3}= -{u}_{d} \cdot {u} + \cos(\eta) \|{u}_{d}\| \| {u} \| \leq 0,
\end{equation}
where $u$ is the actual thrust, along the negative direction of the $\hat{k}_{\B}$ axis, i.e., $\hat{u}=-\hat{k}_{\B}$. To impose \eqref{eqDefCtrlDirectionConst}, an on/off function is applied, which leads to a faster response~\cite{cotorruelo2019output}, rather than handling \eqref{eqDefCtrlDirectionConst} using TSG. Such an on/off method prevents the Deputy spacecraft from applying thrust in a wrong direction: \vspace{-0.05 in}
\begin{equation} \label{eqOnOff}
    u(\tau) := \begin{cases}
         -|u_{d}(\tau)|\hat{k}_{\B}, &\text{if } \pi-\angle u_{d}(\tau)\hat{k}_{\B} \leq \eta , \\
    0, &\text{if }  \pi-\angle u_{d}(\tau)\hat{k}_{\B} > \eta. \end{cases}
\end{equation}  
As \eqref{eqOnOff} is used to enforce \eqref{eqDefCtrlDirectionConst}, the TSG must account for \eqref{eqOnOff} being applied in prediction.

When the Deputy spacecraft operates near the Chief spacecraft, a constraint on the approach velocity is enforced to avoid high-speed collisions. This constraint is activated only when the Deputy spacecraft is in close proximity to the Chief spacecraft, i.e., $\big\| p(X_{d} - X_{c}) \big\| \leq \gamma_{1}$. In such a case, the approach velocity is constrained by a linearly decreasing function of the relative distance from the Deputy to the Chief location, \vspace{-0.05 in}
\begin{equation}
    h_{4} = \big \| v(X_{d} - X_{c}) \big\| - \gamma_{2} \big\|p(X_{d} - X_{c})\big\| - \gamma_{3} \leq 0,
    \label{eqDefh3}
\end{equation}
where $\gamma_{2}$ and $\gamma_{3}$ are constant coefficients.

The convergence of the predicted closed-loop trajectory is restricted to a sufficiently small neighborhood of the target reference at the end of the prediction horizon,
\begin{equation} \label{eqTerminalConstraint}
    h_{5}=\| X_{d}(\tau + \tau_{\tt pred}) - X_{v}(\tau + \tau_{\tt pred}) \| -\epsilon \leq 0 ,
\end{equation}
where $\tau_{\tt pred}$ and $\epsilon$ denote the prediction horizon and the radius of the sufficiently small ball, respectively. Note that this ball must be within the region of attraction of the closed-loop system. We refer to this constraint as the terminal stability constraint. By enforcing~\eqref{eqTerminalConstraint}, the TSG can expand the closed-loop region of attraction.
\vspace{-0.05 in}

\section{TIME SHIFT GOVERNOR} \label{sec:TSG}
\vspace{-0.05 in}
We apply the time shift governor (TSG) to enforce the constraints in a halo orbit rendezvous and docking (RVD) problem in the BCR4BP setting. The TSG augments a nominal closed-loop system consisting of spacecraft dynamics, the ALQR translational controller, and the geometric attitude tracking controller. If there are no constraints (and assuming closed-loop stability), the execution of the RVD with the Chief spacecraft becomes straightforward. In this scenario, the state trajectory of the Chief spacecraft along the reference NRHO is simply governed by the nominal closed-loop system of the Deputy spacecraft.

To avoid constraint violation, the TSG provides both the time shift $\tau_{\tt lead}$ and the time shifted state trajectory of the Chief spacecraft as the reference to the nominal controller of the Deputy spacecraft as \vspace{-0.05 in}
\begin{equation}
    X_{v}(\tau) = X_{c}(\tau + \tau_{\tt lead}),
    \label{eqDefTarget}
\end{equation}
where $\tau_{\tt lead}$ is the time shift. When the Deputy spacecraft is located in front of the Chief spacecraft along the orbital track, the lower and upper bounds of the time shift can be set, respectively, to zero and an initial admissible time shift is $\tau_{\tt lead,0}\geq 0$.

The TSG selects the minimum value of $\tau_{\tt lead} \geq 0$ for which the predicted response over a sufficiently long prediction horizon satisfies the constraints. The update of $\tau_{\tt lead}$ occurs at discrete time instants, and the prediction horizon is chosen sufficiently long to ensure recursive feasibility of the previously chosen value of $\tau_{\tt lead}$. We refer to \cite{kim2023time} for details.

To determine $\tau_{\text{lead}}$, bisections are used. In this process, a time shift candidate $\tau_{\tt lead,m}$ is computed as  \vspace{-0.1 in}
\begin{equation} \label{eqBisection}
    \tau_{\tt lead,m}=f_{\tt mean}(\overline{\tau}_{\tt lead},\underline{\tau}_{\tt lead})=(\overline{\tau}_{\tt lead}+\underline{\tau}_{\tt lead})/2,
\end{equation}
its feasibility is evaluated for the predicted trajectory, and this process repeats until the minimum feasible time shift is determined. We first determine the initial time shift parameter $\tau_{\tt lead, 0} \in \R_{\geq 0}$, such that \eqref{eqNominalCtrl}, \eqref{eqAttControl}, \eqref{eqActualCtrl}, and \eqref{eqDefTarget} with $\tau_{\tt lead} = \tau_{\tt lead, 0}$ result in trajectories satisfying the constraints. In \eqref{eqDefTarget}, we restrict the time shift parameter $\tau_{\tt lead}$ to non-negative values, $\underline\tau_{\tt lead}=0$, with upper bound, $\overline{\tau}_{\tt lead}=\tau_{\tt lead,0}$, i.e., $\tau_{\tt lead}(\tau) \in \Tau = \{ \tau \in \R_{\geq 0}: \underline\tau_{\tt lead}\leq \tau \leq \overline{\tau}_{\tt lead} \},$ 
where $\Tau$ stands for the time shift parameter set. The initial time shift parameter set $\Tau_{\tt 0}$, determined by zero and $\tau_{\tt lead,0}$, is used as an initial guess to search for the next time shift.

Within $\Tau_{\tt 0}$, the TSG iteratively searches for the minimal feasible time shift parameter until the difference between the upper and the lower bounds of $\Tau$ converges to a sufficiently small value. The prediction function evaluates the feasibility of a proposed time shift $\tau_{\tt lead,m} \in \Tau$ in \eqref{eqBisection} for all time instants within a fixed prediction horizon $\tau_{\tt pred}$ based on the current time instant $\tau$, the Chief spacecraft state $X_c(\tau_k)$, and the Deputy spacecraft state $X_d(\tau_k)$. The prediction is based on the forward propagation of the nonlinear model defined in \eqref{eqDyn}, \eqref{eqNominalCtrl}, \eqref{eqSat}, \eqref{eqActualCtrl}, and \eqref{eqOnOff}. Considering \eqref{eqOnOff}, \eqref{eqDefConeConst}, \eqref{eqDefCtrlConst}, \eqref{eqDefCtrlDirectionConst}, and \eqref{eqDefh3}, if the resulting trajectory satisfies the constraints over the prediction horizon, i.e., $\forall \tau \in [\tau_k, \tau_k+\tau_{\tt pred}]$, the time shift candidate $\tau_{\tt lead,m}$ updates the upper bound $\bar{\tau}_{\tt lead}(\tau)$; otherwise the time shift candidate $\tau_{\tt lead,m}$ updates the lower bound $\underline{\tau}_{\tt lead}$. Conversely, it returns zero in the case of any constraint violations. Note the time shift parameter ensures constraint satisfaction for a sufficiently long prediction horizon, i.e., $ \text{$\tau_{\tt pred}$} \gg \tt P_{\tt lead}$, where $\tau_{\tt pred}$ and $\tt P_{\tt lead}$ denote the prediction horizon and the TSG update period. 

After selecting the minimum feasible time shift parameter, the TSG updates the previous $\tau_{\tt lead}$ value with this value. Subsequently, the Deputy spacecraft is chasing the virtual target associated with the current $\tau_{\tt lead}$. At the beginning of the next iteration, the selected minimum time shift parameter $\tau_{\tt lead}(\tau)$ replaces the upper bound $\overline{\tau}_{\tt lead}$ of $\mathcal{T}$ at the time instant $\tau+\tt P_{\tt lead}$, i.e., $\overline{\tau}_{\tt lead}(\tau+\tt P_{\tt lead})=\tau_{\tt lead}(\tau)$. As $\Tau$ updates with $\tau_{\tt lead}$, $\Tau$ gradually shrinks, ensuring the feasible $\tau_{\tt lead}\in \Tau$ becomes a sufficiently small value, which means the Deputy spacecraft achieves the Chief spacecraft without constraint violation if the Deputy spacecraft reaches the virtual target. This TSG update process repeats every $\tt P_{\tt lead}$ until the end of the simulation.

\vspace{-0.05 in}

\section{SIMULATION RESULTS} \label{sec:results}
\vspace{-0.05 in}
Numerical simulations demonstrate the effectiveness of the TSG in enforcing constraints during rendezvous and docking (RVD) missions with the coupled orbit-attitude dynamics model. \vspace{-0.1 in}
\subsection{Simulation Specifications} 
\begin{figure}[htbp!]
    \centering   
  \subfloat{%
        \includegraphics[width=.22 \textwidth]{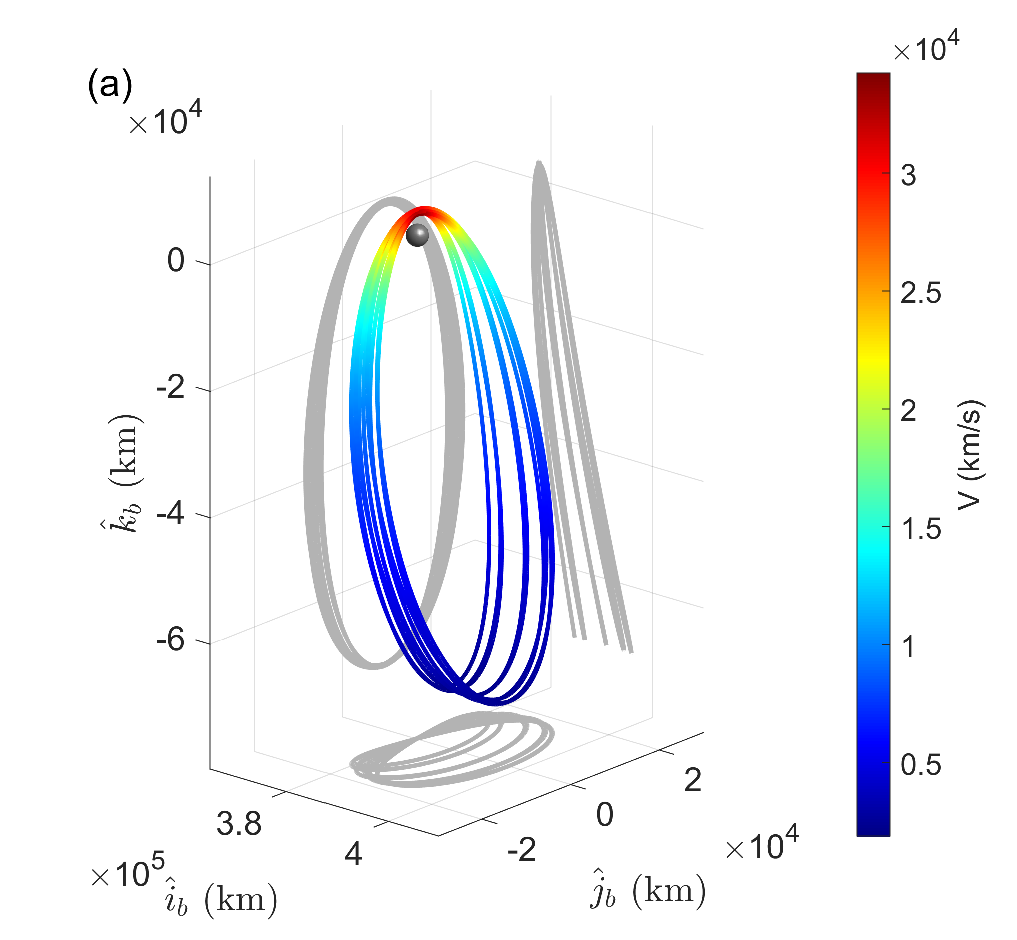} 
        } 
    \hspace{0.03em}%
  \subfloat{%
        \includegraphics[width=.22 \textwidth]{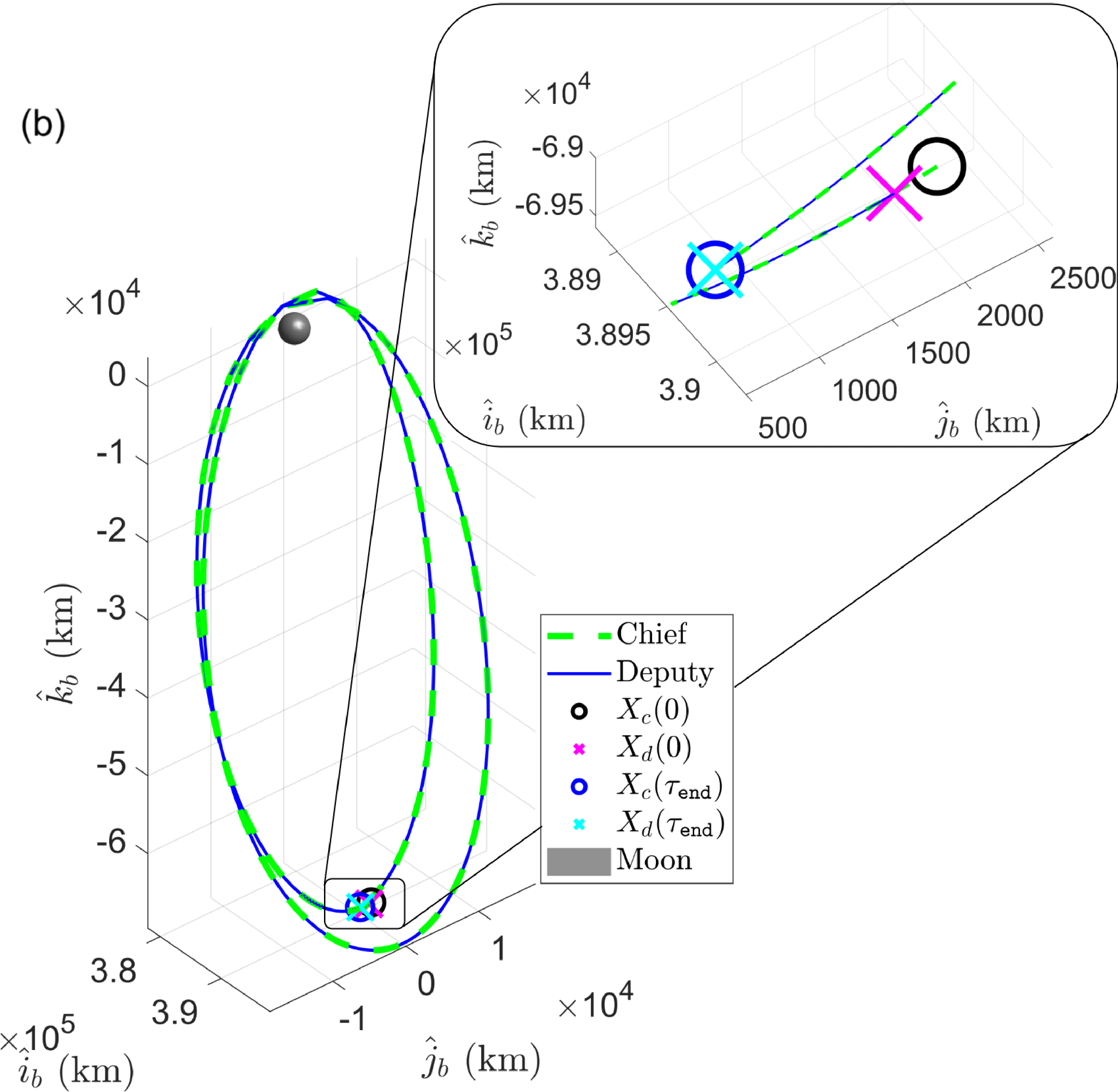} 
        }  \vspace{-0.05 in} 
  \caption{(a) The reference trajectory of nine orbit periods for the Chief spacecraft. (b) the resulting trajectories during the RVD mission with initial and final states.}
  \label{figEntireTraj}    
\end{figure}

\vspace{-0.05 in}
The 9:2 southern $L_2$ NRHO in the Sun-Earth-Moon system is selected as a reference orbit of the Chief spacecraft. Figure~\ref{figEntireTraj}b illustrates the trajectories of the Deputy spacecraft and the Chief spacecraft for two orbit periods. The initial state of the Chief spacecraft, provided from \cite{jpl_periodic_orbits}, is corrected using a shooting method. The initial state of the Deputy spacecraft is chosen as 300 km apart from the initial Chief spacecraft state $X_{c}(0)$ in the reference trajectory. We use the values of mass ratio, length unit, time unit, and moon radius from~\cite{jpl_periodic_orbits}. The prediction horizon $\tau_{\tt pred}$ is selected as 6.56 days corresponding to one orbit period of the reference orbit. 

The nominal closed-loop system of the Deputy spacecraft incorporates translational and rotational controllers. The continuous ALQR controller \eqref{eqNominalCtrl} uses the averaged-in-time LQR gain $K$, associated with the state weight matrix $ Q=\text{diag}(10^6, 10^6, 10^6, 10^3, 10^3, 10^3)$, control weight matrix $\mathcal{R}=\text{diag}(10, 10, 10)$, and the linearized dynamics \eqref{eqLinearizedDyn2Target} for the entire simulation time. Then, the desired control input is computed using \eqref{eqNominalCtrl} and~\eqref{eqSat}. The inertia tensor of the Deputy spacecraft is structured as $[{\tt I_{sc}}]=\text{diag}(4500,\; 4500,\; 1500)\text{ kg}\cdot\text{m}^2$. To align the nozzle direction with the desired, \eqref{eqAttControl} provides the corresponding control torque, and the actual thrust is determined by \eqref{eqOnOff} and \eqref{eqActualCtrl}.   

The imposed constraints, defined in Section~\ref{sec:constraints}, are structured with the following coefficients: The LoS half-cone angle $\alpha$ is 20 deg, the maximum magnitude of the control input $u_{max}$ is $8.2\times10^{-8}\text{ km}\cdot\text{s}^{-2}$, the maximum nozzle angle deviation $\eta$ is 9 deg, and the relative velocity constraint is activated within $\gamma_1=10\text{ km}$ and is designed with $\gamma_2=5.3\times 10^{-5}\text{ s}^{-1}$ and $\gamma_3 =1.0\times 10^{-3}\text{ km}\cdot \text{s}^{-1}$.  We skip the terminal stability constraint $h_{5}$ in \eqref{eqTerminalConstraint} as enforcing it adds to the simulation time.\vspace{-0.05 in}

\subsection{Results} \vspace{-0.05 in}
Crosses mark the initial (magenta) and final (cyan) positions of the Chief spacecraft, while circles mark those of the Deputy spacecraft (initial in black and final in blue). At the end of the simulation, the Deputy spacecraft achieves close proximity to the Chief spacecraft, with a final distance of 6.899 m and a relative velocity of 0.0056 mm/sec.

\vspace{-0.15 in}
 \vspace{-0.15 in}
\vspace{-0.15 in}
\begin{figure}[htbp!]
    \centering
    \includegraphics[width=.21 \textwidth]{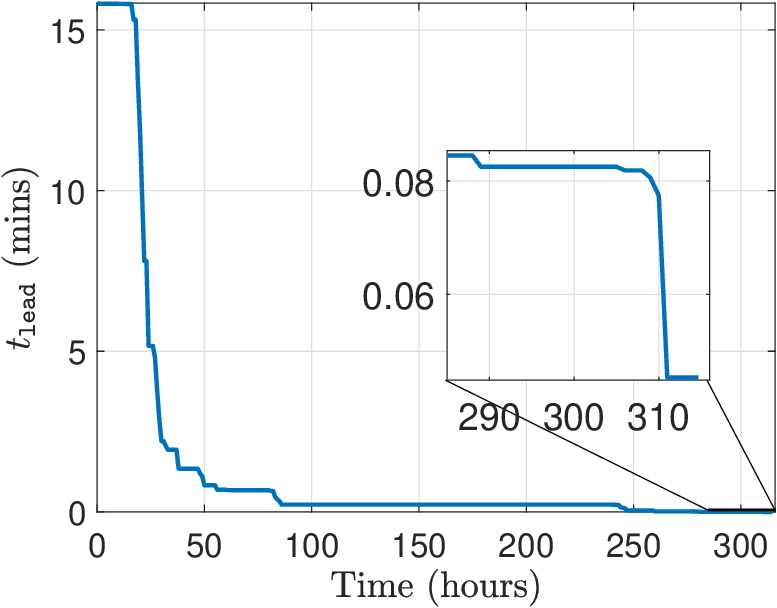} 
\caption{The time shift parameter as a function of time during the RVD scenario. The time unit is dimensionalized by dividing it by the mean motion $n$, i.e., $t_{\tt lead}=\tau_{\tt lead}/n$.}
\label{figtlead}    
\end{figure}
\vspace{-0.15 in}
Figure~\ref{figtlead} shows the evolution of the time shift parameter during the RVD simulation. Initially set to 15.828 min, this parameter is updated hourly and successfully converges to zero at the end.

\begin{figure}[H]
    \centering
    \subfloat{%
       \includegraphics[width=0.46\linewidth]{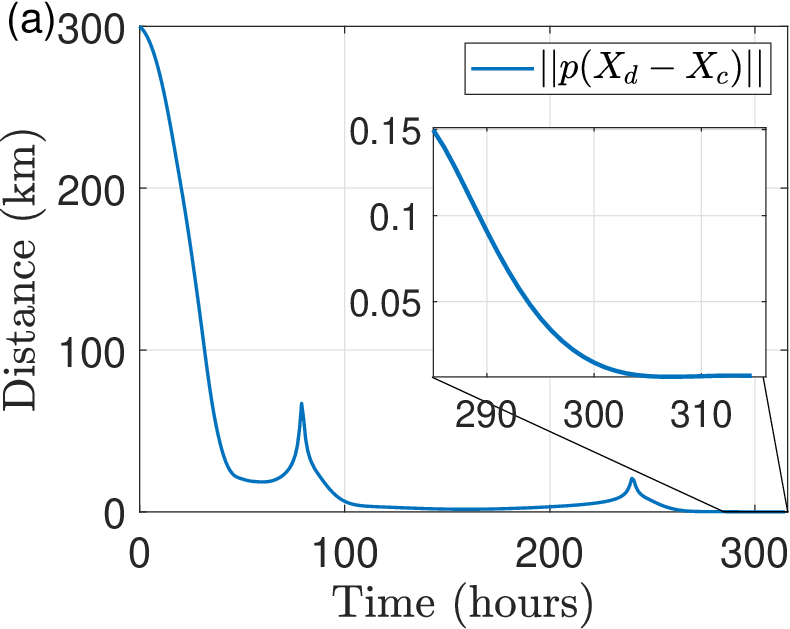}
       } 
    \hspace{0.03em}%
    \subfloat{%
        \includegraphics[width=0.46\linewidth]{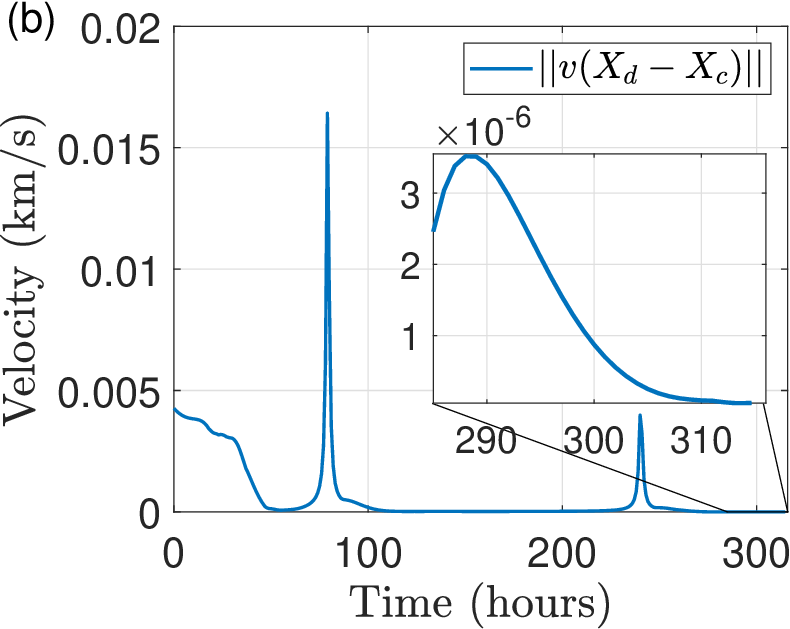}
        } \\ 
    \vspace{-0.1 in}   
    \subfloat{%
        \includegraphics[width=0.46\linewidth]{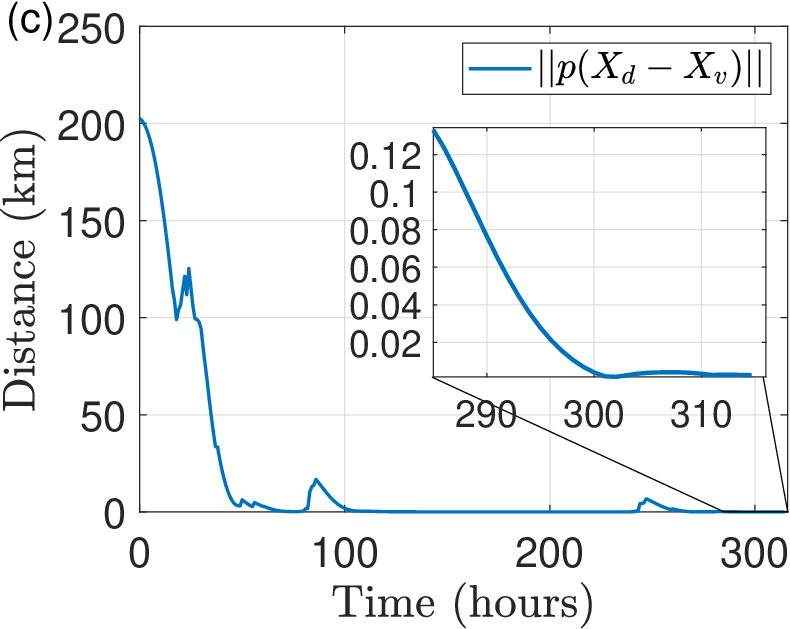}
        }
    \subfloat{%
    \includegraphics[width=0.46\linewidth]{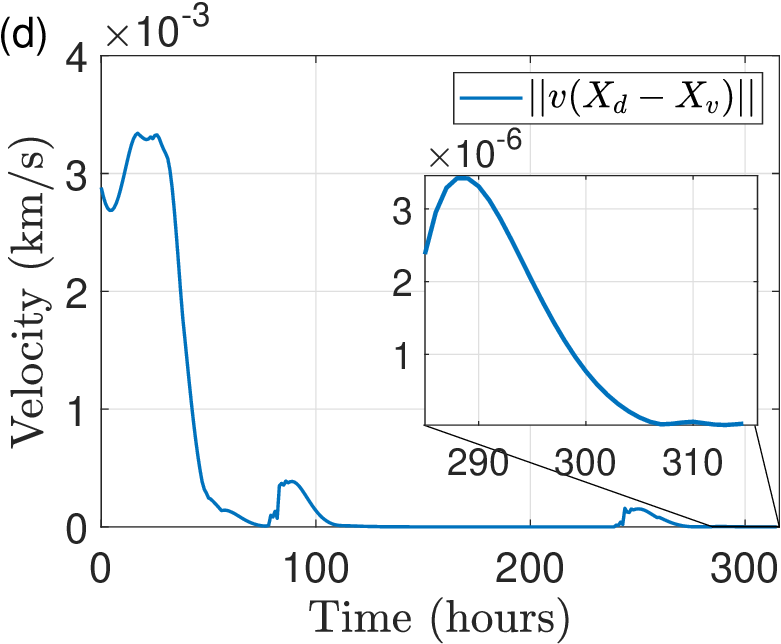}
    }  \vspace{-0.05 in}
  \caption{(a) The relative position and (b) relative velocity of the Deputy spacecraft $X_{d}$ to the Chief spacecraft $X_{c}$. (c) The relative position and (d) relative velocity of $X_{d}$ to the virtual target $X_{v}$.}
  \label{figRelInfo}
\end{figure}

Figure~\ref{figRelInfo} shows the magnitude of the relative position and velocity of the Deputy spacecraft with respect to the Chief spacecraft and the virtual target, respectively. In Figures~\ref{figRelInfo}a and \ref{figRelInfo}b, the Deputy spacecraft achieves close proximity to the Chief spacecraft with a small relative velocity and completes the RVD mission. Figures~\ref{figRelInfo}c and \ref{figRelInfo}d show a relative motion of the Deputy spacecraft with respect to the virtual target. Note less extreme peaks around 80 hours and 240 hours observed in the relative position and velocity presented in Figures~\ref{figRelInfo}c and \ref{figRelInfo}d versus Figures~\ref{figRelInfo}a and \ref{figRelInfo}b. Note that the two spacecraft pass the peri-lunar region approximately every 160 hours, corresponding to the reference orbit period.

\begin{figure}[H]
    \centering
    \subfloat{%
       \includegraphics[width=0.46\linewidth]{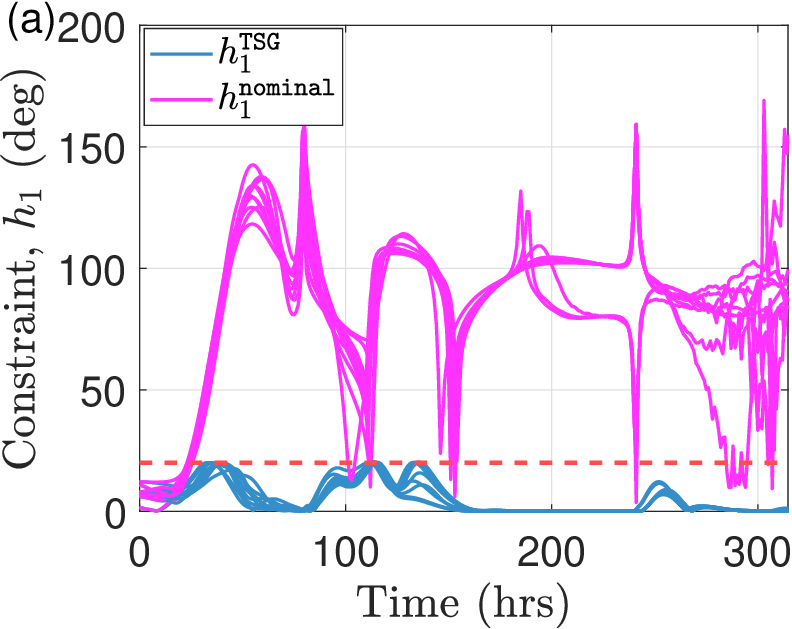}
       } 
    \hspace{0.03em}%
    \subfloat{%
        \includegraphics[width=0.46\linewidth]{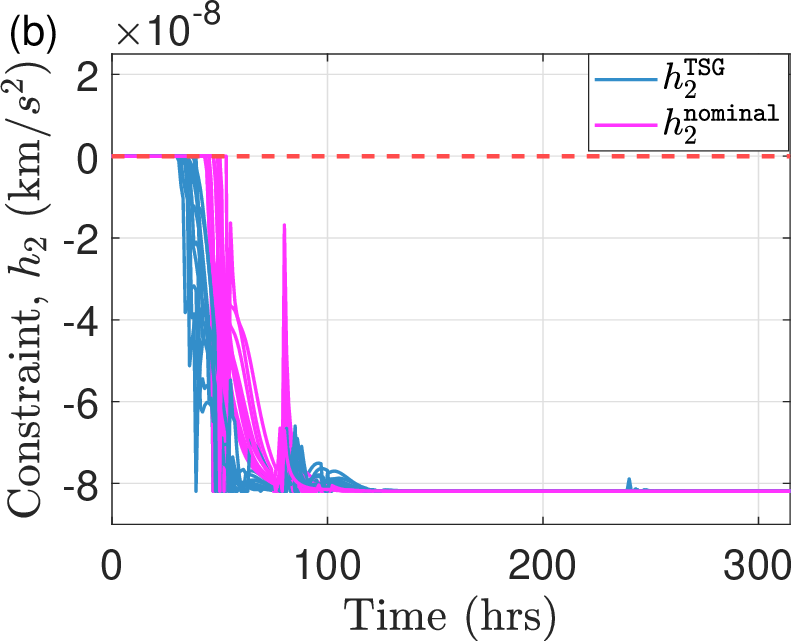}
        } \\ 
    \vspace{-0.1 in}   
    \subfloat{%
        \includegraphics[width=0.46\linewidth]{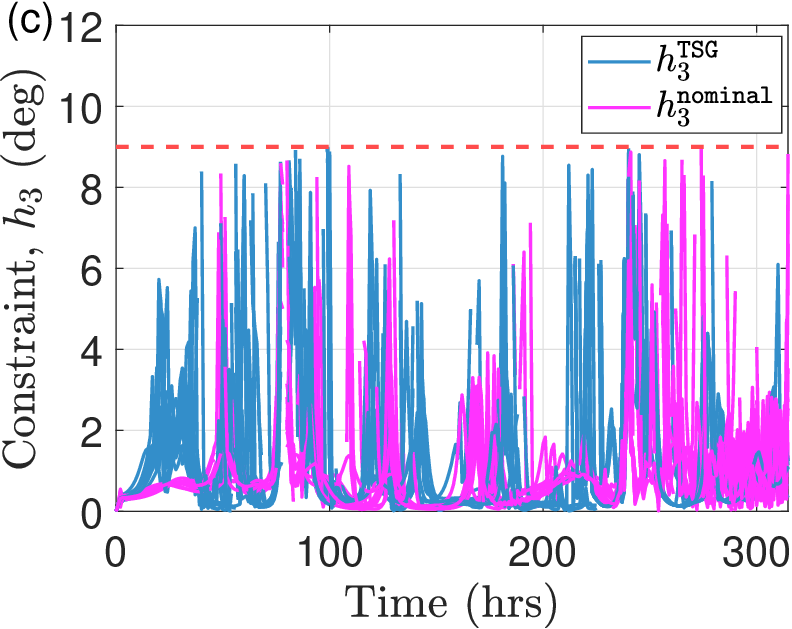}
        }
    \subfloat{%
    \includegraphics[width=0.46\linewidth]{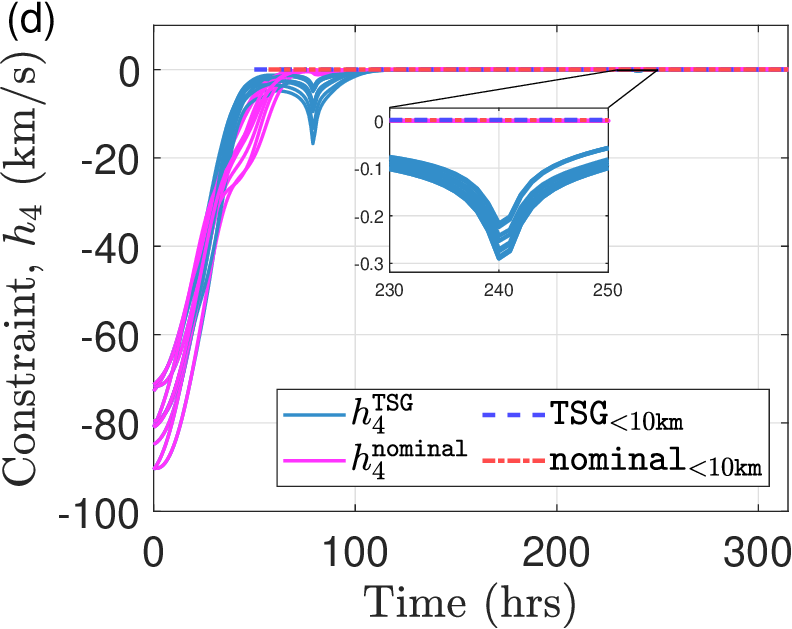}
    }  \vspace{-0.05 in}
  \caption{The constraint trajectories during the RVD simulation using the TSG starting from 10 different initial Deputy states: (a) The LoS cone constraint $h_{1}$; (b) thrust limit $h_{2}$; (c) thrust direction limit $h_{3}$; (d) relative velocity constraint $h_{4}$. Note that the relative velocity constraint activates when the Deputy spacecraft is within 10 km of the Chief spacecraft.}
  \label{figConst}
\end{figure}

Figure~\ref{figConst} illustrates the response of the closed-loop system with and without TSG for ten initial states of the Deputy spacecraft selected as random perturbations in position and velocity also satisfying constraints at the initial time.
demonstrates the effectiveness of our proposed nominal controller with the TSG, compared to when using only the nominal closed-loop system without TSG. Moreover, the robustness of our proposed nominal controller with the TSG is evidenced by the resulting constraint trajectories from the various initial states of the Deputy spacecraft, which include random perturbations in position and velocity and satisfy the constraints at the initial time. Figure~\ref{figConst}a shows the time history for the LoS cone constraint $h_{1}$. In all scenarios, the TSG successfully enforces the LoS cone constraint, while the nominal controller without TSG results in constraint violations.
\vspace{-0.1 in}
\vspace{-0.15 in}
\vspace{-0.18 in}
\vspace{-0.21 in}
Figures~\ref{figConst}b and \ref{figConst}c show the thrust limit $h_{2}$ and the thrust direction limit $h_{3}$, respectively, and these constraints are satisfied. Furthermore, Figure~\ref{figConst}b shows that using the TSG leads to less amount of required control input over the RVD mission, given our proposed nominal controller.

Figure~\ref{figConst}d shows the time history of the approach velocity constraint $h_{4}$. The nominal controller alone and the nominal controller augmented by the TSG satisfy this constraint. However, the Deputy spacecraft approaches the Chief spacecraft in a direction other than the docking port direction, leading to a collision. This failure is due to the LoS constraint being satisfied only during the initial 24 hours, while the Deputy remains more than 10 km away from the Chief spacecraft.
\vspace{-0.15 in}
\begin{figure}[htbp!]
    \centering   
    \subfloat{%
        \includegraphics[width=0.46\linewidth]{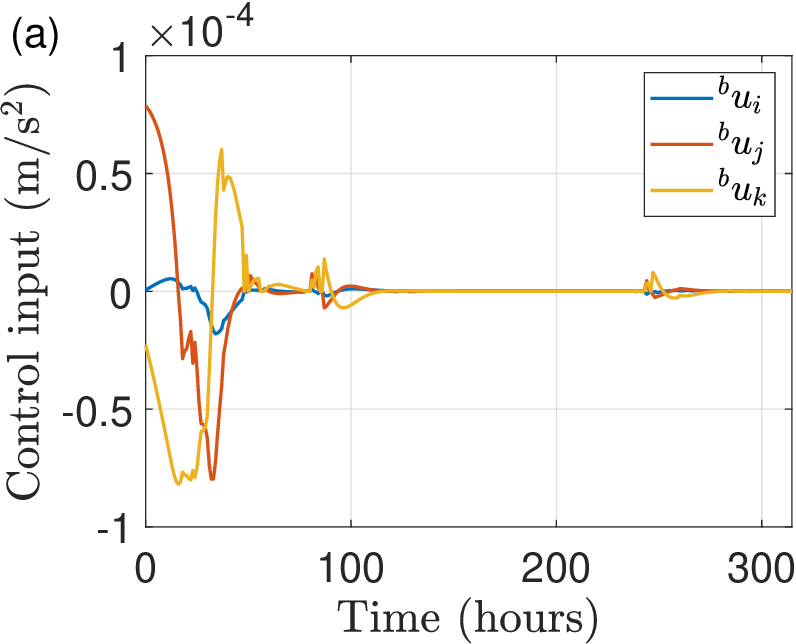}
        } 
    \hspace{0.5em}%
  \subfloat{%
        \includegraphics[width=0.46\linewidth]{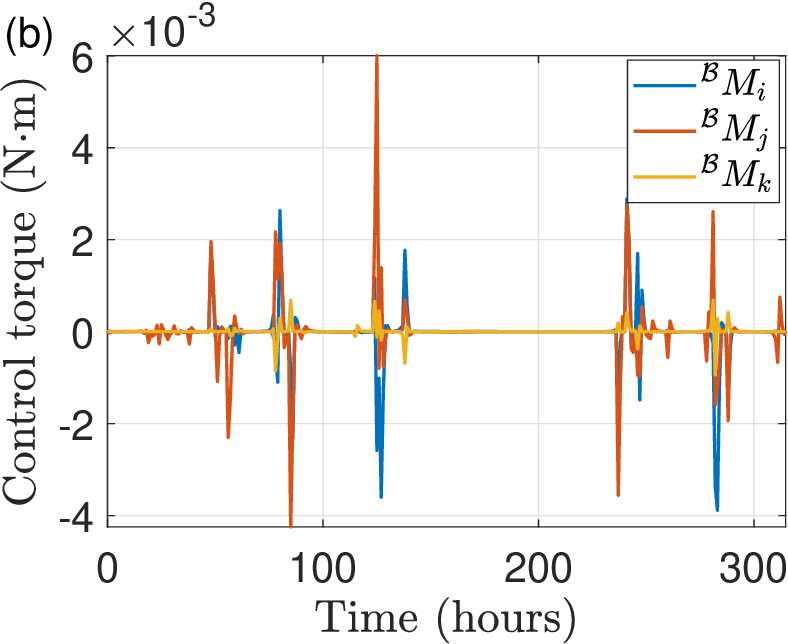}
        }  \vspace{-0.05 in} 
  \caption{(a) The time histories of the desired control input, $u_{d}$, expressed in the barycentric frame $b$. (b) The applied control torque histories, expressed in the body-fixed frame $\B$.}
  \label{figControl}
\end{figure} \vspace{-0.15 in}
\vspace{-0.2 in}
\begin{figure}[htbp!]
    \centering   
  \subfloat{%
        \includegraphics[width=0.46\linewidth]{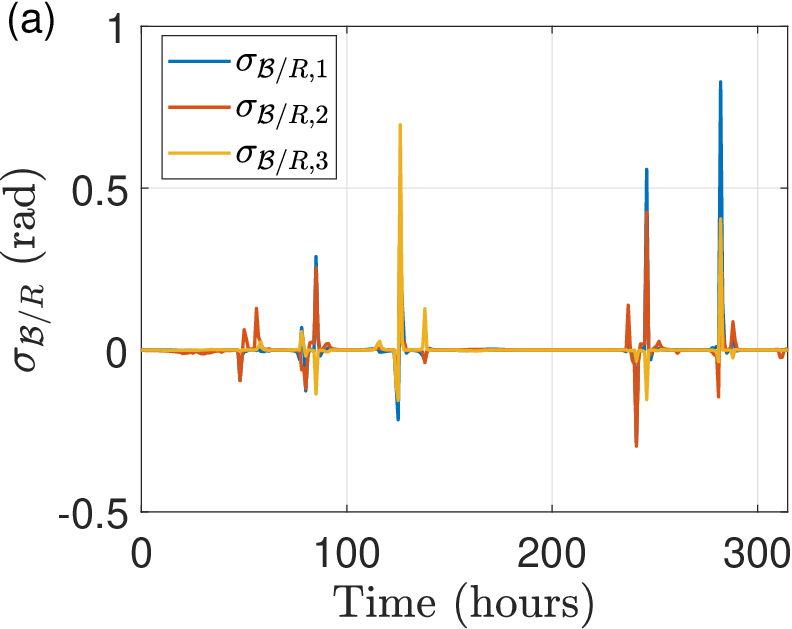}
        } 
    \hspace{0.5em}%
  \subfloat{%
        \includegraphics[width=0.46\linewidth]{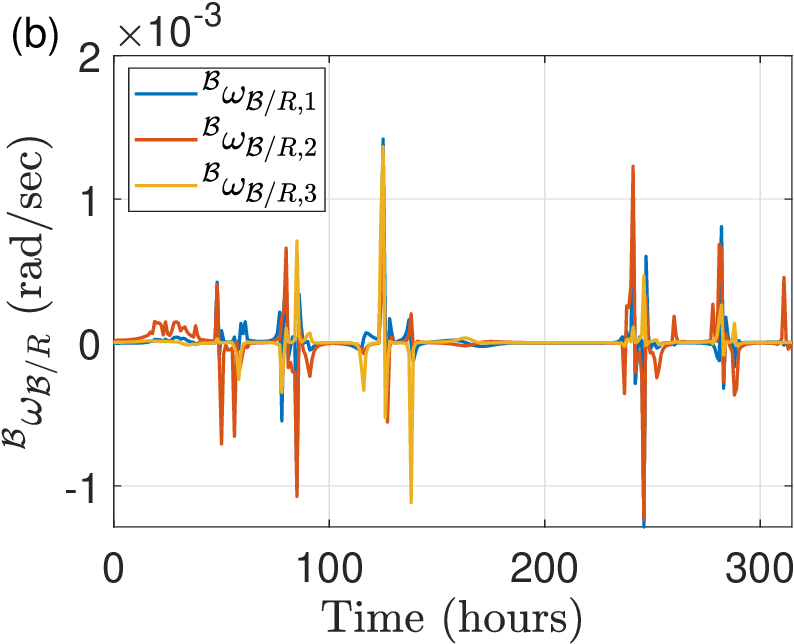}
        }  \vspace{-0.05 in} 
 \caption{(a) The attitude of the body-fixed frame $\B$ expressed in MRPs, relative to the desired reference attitude $R$ in \eqref{eqDesiredAtt}, and (b) components of angular velocity vector expressed in the body-fixed frame $\B$.}
  \label{figAtt}
\end{figure} \vspace{-0.15 in}

Figure~\ref{figControl}a presents the desired control inputs, which remain near zero after 110 hours. Figure~\ref{figControl}b presents the applied control torque expressed in the body-fixed frame $\B$ used to align the spacecraft with the desired thrust direction.

Figure~\ref{figAtt} displays the trajectories of the Deputy spacecraft's attitude and angular velocity with respect to the reference attitude~\eqref{eqDesiredAtt}. Figure~\ref{figAtt}a demonstrates that the actual thrust generally aligns with the desired control input, with minor exceptions. This indicates that the geometric tracking controller in \eqref{eqAttControl} has successfully stabilized the spacecraft attitude, enabling the Deputy spacecraft to track its target with the ALQR controller in \eqref{eqNominalCtrl}, as seen in Figures~\ref{figRelInfo}c and \ref{figRelInfo}d.
\vspace{-0.05 in}

\section{CONCLUSIONS} \label{sec:conclusion}
\vspace{-0.05 in}
We presented a coupled orbit and attitude dynamic model for a spacecraft in the Bicircular Restricted Four-Body Problem setting. We also proposed a nominal control system for tracking translational and rotational motions. We developed the Time Shift Governor (TSG) to handle constraints during spacecraft rendezvous and docking (RVD) missions in a Near Rectilinear Halo Orbit (NRHO). The TSG has demonstrated its ability to enforce multiple constraints during RVD simulations, including the line of sight cone constraint, thrust limit, thrust direction limit, and relative velocity constraint. Over time, the time shift parameter of the TSG converges to zero, aligning the virtual target and the Chief spacecraft. Simulated maneuvers in an NRHO within the Sun-Earth-Moon system have confirmed the effectiveness of the TSG in addressing the constraints. 

\vspace{-0.05 in}


\addtolength{\textheight}{-12cm}   





\bibliographystyle{IEEEtran}

\bibliography{references}


\end{document}